\documentclass[letterpaper,twocolumn,english,aps,prl,floatfix,showpacs]{revtex4}
\usepackage{amsmath}
\usepackage{graphicx}
\usepackage{amssymb}

\makeatletter
\usepackage{bm}

\makeatother
\begin{document}

\title{Robust $d$-wave Pairing Correlations in the Heisenberg Kondo Lattice
Model }

\author{J.~C.~Xavier}

\affiliation{Instituto de F\'{\i}sica, Universidade Federal de Uberl\^andia, C.P.
593, Uberl\^andia MG 38400-902, Brazil}

\author{E.~Dagotto}

\affiliation{ Department of Physics, University of Tennessee, Knoxville, TN 37996, and \\
Materials Science and Technology Division, Oak Ridge National Laboratory, Oak Ridge, TN 37831}

\begin{abstract}
The Kondo lattice model enlarged by an antiferromagnetic coupling $J_{\rm AF}$
between the localized spins is here investigated using computational techniques.
Our results suggest the existence of a $d$-wave superconducting phase
close to half-filling mediated by antiferromagnetic fluctuations.
This establishes a closer
connection between theory and heavy fermion experiments 
than currently
provided by the standard Kondo lattice model with $J_{\rm AF}$=0.
\end{abstract}


\date{\today{}}

\maketitle
{\it Introduction.}
Heavy fermions (HF) continue attracting the interest of the condensed
matter community \cite{reviewHF}. These materials are 
phenomenologically described by Doniach's scenario, where localized
spins interact with conduction electrons via an exchance interaction $J$ \cite{Doniach}.
At small $J$, the indirect Ruderman-Kittel-Kasuya-Yosida (RKKY) mechanism
is expected to induce an antiferromagnetic (AF) state, whereas for large $J$ a paramagnetic
spin-liquid state emerges \cite{Doniach}. The discovery
of superconductivity in $\rm CeCu_{2}Si_{2}$ \cite{firsthfsuper}, and subsequently
in several other HF compounds,
unveiled the rich variety of phenomena present in these
strongly correlated electronic systems. Currently, HF superconductors 
are widely considered as examples of superconductivity not mediated 
by lattice vibrations \cite{mathuretal,jourdan}.

Several formal theories of HF materials start 
with the Kondo lattice model (KLM) \cite{hewson}. 
The development of powerful many-body numerical techniques,
and the continuous growth in computer power, have allowed for 
non mean-field and free of non-controlled parameters 
investigations of the KLM, at least in low dimensional systems. 
These investigations have revealed two potentially important problems in
establishing a connection between KLM and HF phenomenology: 
(1) Including hole carriers away
from half-filling, and at large $J$, the KLM leads to a robust ferromagnetic state that
is not obviously connected with states found experimentally \cite{hiro}; 
(2) more importantly,
there is evidence that the standard one-dimensional (1D) KLM does $not$ present
SC tendencies
close to half-filling \cite{xavierqcp}. Although these anomalies
may be caused by the low dimensionality of the Kondo lattices 
studied, 
they still raise doubts about the full validity of the simple KLM to describe 
heavy fermions.
Moreover, it would be advantageous for theoretical investigations to identify 
a simple HF model with clear SC tendencies, {\it 
even in low dimensions}.
The growing number of SC HF compounds clearly requires a 
model paradigm involving
not only AF and spin-gapped phases, as in the past, 
but including a SC phase as well.

To address these concerns, and better capture the physics of HF systems, we
must move beyond the standard KLM. In this letter, using computational
techniques it is shown that the addition
of a direct AF coupling $J_{\rm AF}$ between the localized spins 
considerably alleviates the problems mentioned above. 
Previous investigations had already suggested that for 
$\rm UPd_{2}Al_{3}$ the AF coupling between
localized spins is important to understand the
SC phase \cite{satoetal}. 
Moreover, other authors already remarked the importance of $J_{\rm AF}$ when
focusing on the magnetism of the HF systems
\cite{xaviermal,dmrg-affleck-white,moukouri}. However, our present effort goes 
beyond these
previous studies by revealing
the formation of $d$-wave symmetric hole-pairs and the
development of robust SC correlations when $J_{\rm AF}$ is incorporated to the 
two-dimensional  KLM.
This establishes a better connection theory-experiment 
in the HF context than provided by the plain KLM.

{\it Model/Methods.} 
The antiferromagnetic Heisenberg Kondo Lattice Model (HKLM) 
is given  by
\[
H=-\sum_{\langle i,j \rangle ,\sigma}(c_{i,\sigma}^{\dagger}c_{j,\sigma}^{\phantom{\dagger}}+\mathrm{H.}\,\mathrm{c.})+J\sum_{j}\mathbf{S}_{j}\cdot\mathbf{s}_{j} +J_{\rm AF}\sum_{\langle i,j \rangle}\mathbf{S}_{i}\cdot\mathbf{S}_{j},\]
\noindent 
where $J$$>$$0$
is the Kondo coupling between the conduction electrons and
the local moments, $J_{\rm AF}$ is the antiferromagnetic interaction
between the localized spins-1/2, and the hopping amplitude was set to
unity to fix the energy scale. The rest of the notation is standard.
The HKLM 
on a $N$$\times$$L$
cluster  was here investigated using the Lanczos technique \cite{dagottorev},
and the DMRG method \cite{white} under open boundary conditions (OBC)
\cite{dmrg-previous,comment0}.

Note that the same HKLM also describes
the manganites if $J$$<$$0$ and the $\mathbf{S}_{j}$'s  
are assumed classical \cite{reviewElbio}. 
Investigations of manganite models have shown that 
$J_{\rm AF}$ is crucial
for the numerical stabilization of
experimentally known phases that otherwise become 
unstable due to the strong
ferromagnetic tendencies \cite{reviewElbio,JAFmanga}. Thus, our effort also provides a
unifying view of HF and manganite research regarding the relevance of the $J_{\rm AF}$ coupling.

\begin{figure}[tbp]
\begin{centering}\includegraphics[width=5.2cm,height=3.5cm]{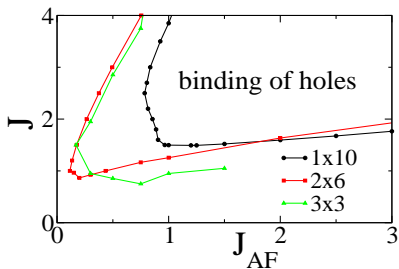}\par\end{centering}

\caption{\label{fig1}(Color online) The region with binding of holes in the $J$-$J_{\rm AF}$ plane, 
for different cluster shapes with OBC. 
The solid line corresponds to $\Delta_{\rm B}$$=$$-0.1.$}
\end{figure}

{\it Binding Energies.}  Our investigations mainly focused on pairing and SC correlations.
To observe indications of pairing in the HKLM ground state,
we measure the pair-binding energy defined as 
$\Delta_{\rm B}$=$E(0)$+$E(2)$-$2E(1)$,
where $E(l)$ is the ground state energy in the subspace with $(NL-l)$ conduction electrons.
If the holes do not form a bound state, in a \emph{finite system} the
binding energy is positive $\Delta_{\rm B}$$>$$0$ \cite{dagottorev}, while
in the thermodynamic limit it vanishes. On the other hand, if holes do
form a bound state then $\Delta_{\rm B}$$<$$0$, and this is indicative that
attractive effective forces are present, as widely discussed before in the
context of high-T$_c$ investigations \cite{dagottorev}. 

In Fig.~1, we show the region in the $J$-$J_{\rm AF}$ plane 
where $\Delta_{\rm B}$ is less than -0.1, 
for the cluster sizes 1$\times$10, 2$\times$6  
and for a square
lattice 3$\times$3 with OBC.
We choose to present the region where $\Delta_{\rm B}$$<$$-0.1$,
as opposed to $\Delta_{\rm B}$=0, since previous experience in the cuprate 
context \cite{dagottorev}
suggests that this procedure takes better into account the size effects. 
The limitation of using small clusters originates in the huge Hilbert
space of the HKLM, with 8 states per site, and in the need
to calculate hundreds of points to extract comprehensive phase diagrams. 
However, for some selected couplings much larger clusters were considered
and a thermodynamic limit extrapolation was carried out
(see Fig.~2 and discussion below). Note that
the binding regions in Fig.~1 are qualitatively 
similar, although the clusters 
have different shapes.
Thus, the formation of hole pairs is a robust effect, suggesting
that pair formation might also appear in the three dimensional phase diagram 
(currently unreachable numerically).
Moreover, the tendency toward increasing pairing strength found in Fig.~1
moving from chains to ladders/square clusters is similar to tendencies
found in the $t$-$J$ model for cuprates \cite{dagottorev}. 
It is interesting also to note that a previous study by Sikkeman {\it et al.}
\cite{dmrg-affleck-white} have found that $J_{\rm AF}$ induces a spin gap phase
with no charge gap close to half-filling for the one-dimensional HKLM, in
agreement with the  phase diagram present in Fig.~1.
Finally, note that  values of $J_{\rm AF}$ are expected to be small due to the small overlap between
nearest-neighbor 4f/5f orbitals, and in  Fig.~1(c) it appears that hole
binding can occur for $J_{\rm AF}$ as small as 0.2. Only
a careful two-dimensional size scaling, beyond our capabilities at present, can answer how
small $J_{\rm AF}$ can be to still induce hole pairing.

In Figs.~2(a)-(c), $\Delta_{\rm B}$ vs. $1/L$
is shown for the chain, two-leg ladder  and square clusters, respectively.
There are clearly two distinct behaviors. 
For the couplings that present
hole binding for the small clusters presented in Figs.~1,
$\Delta_{\rm B}$ tends to negative values when $L$$\rightarrow$$\infty$, 
while for the other parameters $\Delta_{\rm B}$$\rightarrow$0.
Regarding the 6$\times$6 cluster,
even working with up to $m$=$4000$ states the energies did not converge well. 
However, as presented in Fig. 2(d), $\Delta_{\rm B}$ tends to a negative  value 
for the intermediate coupling range, similar as the results
obtained for the other cluster shapes.
All these results indicate that close to
half-filling, pairing tendencies exist for intermediate values of $J$ and
$J_{\rm AF}$ in the bulk of the HKLM.
It is also interesting to observe the behavior of $\Delta_{\rm B}$ vs. 
$J$ for a fixed $J_{\rm AF}$ (or fixing
$J$ and varying $J_{\rm AF}$). As present in Fig. 3(a), 
$-\Delta_{\rm B}$ reaches a maximum value at $intermediate$ $J$,
important detail to compare with experiments, as discussed later.  
We  found that the hole binding is robust
in the same  region where the nearest-neighbor spin-spin correlations
are robust, as shown in the inset of Fig. 3(a).
There, we also plot the nearest-neighbor hole-hole correlation \cite{comment_hole}. The fact
that spin and hole correlations behave similarly supports our claim that antiferromagnetism and hole
binding are related. At very large $J$, both quantities are suppressed together.
This suggest that the origin of the hole-pair attraction is connected with
AF fluctuations. We believe mechanisms similar to those identified in cuprates \cite{dagottorev},
such as hole attraction to minimize the ``damage'' to the AF background, 
are in action in the HKLM as well

\begin{figure}[tbp]
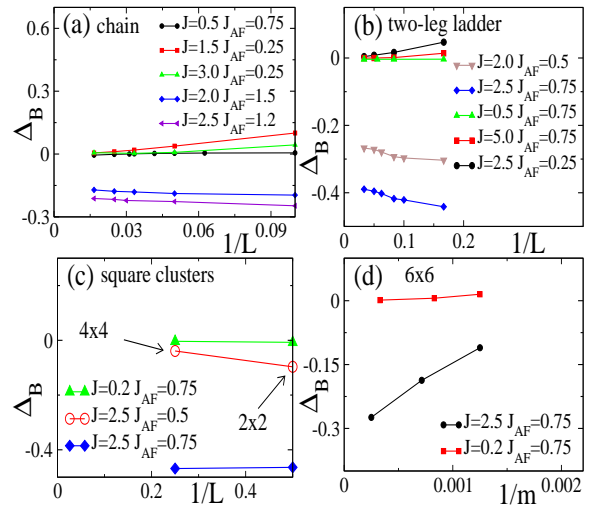

\begin{centering}\includegraphics[width=3.8cm,height=3.3cm]{fig2a}\includegraphics[width=3.8cm,height=3.3cm]{fig2b}\par\end{centering}
\begin{centering}\includegraphics[width=3.8cm,height=3.3cm]{fig2c}\includegraphics[width=3.8cm,height=3.3cm]{fig2d}\par\end{centering}
\caption{\label{fig2}(Color online) Pair-binding energies $\Delta_{\rm B}$
vs. $1/L$ for  a chain (a), a two-leg ladder (b), and square clusters (c). 
(d) $\Delta_{\rm B}$ vs. $1/m$ for the cluster 6$\times$6, at
couplings indicated. }
\end{figure}

{\it Pair Symmetry.}
Our exact diagonalization results for the clusters 2$\times$2 and 3$\times$3 show that the ground
state with 0 (2) holes has $s$-wave ($d$-wave) symmetry under rotations. 
Thus, these ground states are connected via a pair-creation operator 
with $d$-wave symmetry,
similarly as observed in the Hubbard and $t$-$J$
 models \cite{dagottorev}. Then, 
the HKLM predicts $d$-wave superconductivity, compatible with 
previous results for HF systems \cite{jourdan}.

{\it Pairing Correlations.}
To confirm that hole-pair tendencies are concomitant with robust superconducting correlations,
for the two-leg ladder we measured the rung-rung pair correlation functions
$C(l)$=$\left\langle \Delta_{i+l}\Delta_{i}^{\dagger}\right\rangle$
where the pair operator is defined 
as $\Delta^{\dagger}_l$=$c_{l1,\uparrow}^{\dagger}c_{l2,\downarrow}$-$c_{l1,\downarrow}^{\dagger}c_{l2,\uparrow}$, 
and $c_{l\lambda,\sigma}$ annihilates a conduction electron on
rung $l$ and leg $\lambda=1,2$ with spins $\sigma=\uparrow,\downarrow$ \cite{comment2}.
\begin{figure}[tbp]
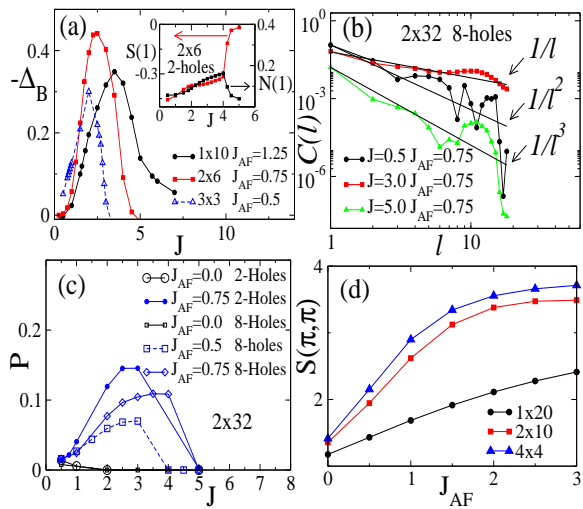

\begin{centering}\includegraphics[width=3.8cm,height=3.3cm]{fig3a}\includegraphics[width=3.8cm,height=3.3cm]{fig3b}\par\end{centering}
\begin{centering}\includegraphics[width=3.8cm,height=3.3cm]{fig3c}\includegraphics[width=3.8cm,height=3.3cm]{fig3d}\par\end{centering}
\caption{\label{fig3} (Color online) (a) $-\Delta_{B}$ vs. $J$
for the clusters 1$\times$10, 2$\times$6, and 3$\times$3 with OBC. Inset: 
The  nearest-neighbor spin-spin  ($S(1)$) and hole-hole ($N(1)$) correlations,  
measured in the center of the 2$\times$6 cluster 
with 2-holes and $J_{\rm AF}$=$0.75$,  
as function of $J$. (b) The
pair-pair correlation function $C(l)$ vs. $l$ for a system size
2$\times$32 and $J_{\rm AF}$=$0.75$ with 8 holes, for some values of $J$. 
The straight lines are data fits with the powers indicated by the arrows.
(c)
$P$ vs. $J$ for the cluster 2$\times$32 with 2 and 8 holes and some values
of $J_{\rm AF}$. (d) The spin structure factor $S(\pi,\pi)$ vs. $J_{\rm AF}$ for
the clusters 1$\times$20, 2$\times$10 and 4$\times$4, at $J$=$2$ 
and half-filling.  }
\end{figure}
In Fig.~3(b), $C(l)$ is shown vs. distance $l$,
for a fixed $J_{\rm AF}$ and some values of $J$, using a cluster 2$\times$32
with 8 holes. Similar results are obtained with 2 holes. In order to obtain
the {\em slope} of $C(l)$ we fit our data with the function  $a/L^n$ ($n$ an integer). 
Clearly from this figure, the pair-pair correlations $C(l)$ are enhanced
at large distances for intermediate coupling values, i. e. $J$=$3$ and 
$J_{\rm AF}$=$0.75$, with a robust power-law decay $\sim$$1/l$.
It is interesting to note that the well known 
two-leg $t$-$J$ model also has the same slope close to half-filling 
\cite{hayward,ladderrev}.
The origin of this effect may be due to the fact that 
the HKLM can be mapped, in a limiting case,  into the  $t$-$J$ model \cite{dmrg-affleck-white}.
Similar SC tendencies are also observed in Fig.~3(c), 
where $P$=$\sum_{l=5}^{18}C(l)$ is shown vs.
$J$, at a fixed $J_{\rm AF}$. 
$P$ is
enhanced $only$ for non-zero values of $J_{\rm AF}$, indicating that superconductivity
appears only when this interaction is active.  
Although there is no true long-range order
in quasi-1D systems, our results provide strong
evidence that superconductivity dominates at intermediate
$J$ and $J_{\rm AF}$ in the HKLM close to half-filling. Similar conclusions
were reached for the $t$-$J$ model on chains and ladders \cite{dagottorev,hayward,ladderrev}.

{\it Magnetic Properties.}
We also investigated 
the effect of $J_{\rm AF}$ in the magnetic properties of the HKLM.
At small values of $J$, due to the  the RKKY interaction, antiferromagnetic 
long-range order  is expected, whereas for large $J$ a paramagnetic
state must emerge. The competition between these two states leads to a quantum
critical point at $J_c\sim1.45$ for the two-dimensional 
KLM at half-filling \cite{Doniach,qcp1,xavierqcp,assaad}. 
If the AF coupling between the localized spins is added, we favor 
antiferromagnetism even more. 
For this reason, $J_c$ is expected to increase 
with $J_{\rm AF}$. Here, we do not attempt to provide the location of
$J_c(J_{\rm AF})$ for the two-dimensional HKLM, which is a formidable task,
but only show the dominant tendencies in the problem.
In Fig.~3(d), we present the intensity of the spin structure factor 
$S(\vec q)$ at $\vec q$=$(\pi,\pi)$ vs. $J_{\rm AF}$, 
for several clusters
at $J$=2.0 and half-filling 
Clearly from this figure, $S(\pi,\pi)$   increases 
with $J_{\rm AF}$, suggesting  that   $J_c$ will increase  as well, as anticipated.
As in the two-leg ladder \cite{twolegpd}, we also have found no evidence of
ferromagnetism close to half-filling for the 4$\times$4 cluster.
Note also that it is expected  that 
the AF-phase will survive away from half-filling,
for the three-dimensional KLM \cite{hiro}. We believe  this phase will
also exist
in the two-dimensional KLM close to half-filling.
 

{\it Discussion.}
Based on the numerical results described in this investigation, in Fig.~4(a) a 
qualitative phase diagram for the HKLM  close to half-filling is presented. 
The robustness of our results with respect to the shape of the cluster used, 
suggest our conclusions may be qualitatively valid even in higher 
dimensions.    
The only mild assumption in Fig.~4 is the existence of an overlap between the AF and
$d$-wave superconducting regions. At a fixed $J$ and with increasing $J_{\rm AF}$ this 
overlap is to be expected. In the SC phase not overlapping with AF, strong
short-range AF fluctuations are also to be expected.

\begin{figure}[tbp]
\begin{centering}\includegraphics[scale=0.27]{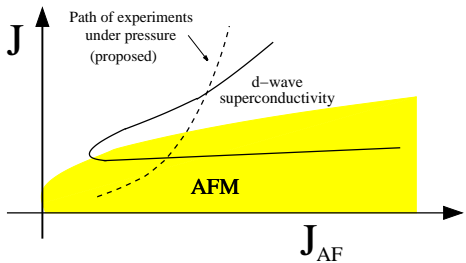}\par\end{centering}
\caption{\label{fig4} (Color online) Schematic phase diagram of the HKLM 
close to half-filling. The solid line defines
the region where $d$-wave SC should exist based on our numerical calculations.
The yellow region is the long-range 
AF phase. The dashed line describes a trajectory 
with increasing pressure that is compatible with HF experiments (see text).
}
\end{figure}

The phase diagram in Fig.~4 is in qualitative agreement with
experimental results reported for some HF materials, such as $\rm CeIn_{3}$ and $\rm CePd_{2}Si_{2}$,
at low temperatures. A possible experimental trajectory
is shown as a dashed line in Fig.~4.
At ambient pressure these compounds are known to be AF \cite{mathuretal}. 
As observed in the figure, for small values of $J$
and $J_{\rm AF}$ (corresponding to low pressures in the experiments)
there is no binding of holes (see Fig.~1) and superconductivity is
not expected. However, with increasing
pressure at a critical value
$p_{c_{1}}$, and within a narrow pressure range, superconductivity develops and
coexists with long-range AF order \cite{mathuretal}. 
This is compatible with our results, since at intermediate values of $J$ and
$J_{\rm AF}$  (intermediate pressures) a tendency to $d$-wave
superconductivity was numerically observed.
At pressures even higher $p_{c_{2}}$$>$$p_{c_{1}}$,
the real HF system first stabilizes a SC state without AF long-range order,
and then finally a transition to a non-SC paramagnetic phase is reached. 
Again, from the theory perspective this is reasonable since 
for even larger values of $J$ and $J_{\rm AF}$ (larger
pressures) the system eventually transitions into a phase without hole binding
(see Figs.~1,2).  Note that the positive slope of the proposed ``path'' in
Fig.~4 is  qualitatively correct 
since under pressure both $J$ and $J_{\rm AF}$ increase,  due to the increasing
overlaps of wave functions. 
Note also that theoretically the maximum binding energy is reached at
intermediate couplings in the dashed-line path, suggesting that the SC critical
temperature first increases under pressure, reaches a maximum, and then
decreases,  as in experiments \cite{mathuretal,comment3}.


Summarizing, we have investigated the HKLM 
with emphasis on hole-pair formation and SC correlations. 
A region with a robust tendency toward a SC state was identified
at intermediate values of $J$ and $J_{\rm AF}$ close to half-filling. The binding of holes only
appears when the AF interaction between the localized
spins is considered, showing its critical importance for a proper theoretical
description of heavy fermion materials.

\begin{acknowledgments}
Research supported by the Brazilian agencies FAPEMIG and CNPq,
NSF grant DMR-0706020, and the Division of Materials Sciences and
Engineering, U.S. DOE under contract with UT-Battelle, LLC. Useful comments by
T. Hotta are acknowledged.

\end{acknowledgments}
\bibliographystyle{apsrev}

\begin{thebibliography}{17}
\bibitem{comment0}
We have done $\sim$10-25
sweeps, and the  discarded weight was typically $10^{-6}$-$10^{-8}$ in that
final sweep for the chains and ladders, while for the square clusters
4$\times$4 and 6$\times$6 it was typically $10^{-4}$-$10^{-6}$.
%
Here, we have defined the estimated error of the energy  
as $\Delta E$=$E(m_{max})-E(m_{max}/2)$, where $E(m_{max})$
is the energy obtained  with $m=m_{max}$ states (we use $m_{max}$ $\sim$800-4000,
depending of the cluster sizes and the
parameters).   In most of our results  
$\Delta E<10^{-4}$, although for a few cases  $\Delta E$ was
$\sim 5\times10^{-2}$. 
In all results presented here, the errors are smaller than 
the sizes of symbols presented in the figures.
\bibitem{comment_hole} We defined the hole-hole correlations as
  $<\bar n(i)\bar n(j)>$, where $\bar n(i)=(1-n_{\uparrow}(i))(1-n_{\downarrow}(i))$.
\bibitem{comment2}
To minimize the influence of the boundary, $C(l)$ was measured in the center
portion of the two-leg ladder.
\bibitem{comment3} Paths with the opposite curvature, e.g. $J$$\sim$$J_{\rm AF}^{1/2}$ as
proposed by Iglesias {\it et al.} \cite{xaviermal}, can also be shown to be
compatible with experiments.
\end{thebibliography}

\end{document}